\documentclass[12pt,a4paper,english]{article}
\usepackage[T1]{fontenc}
\usepackage[latin2]{inputenc}
\usepackage[english]{babel}
\usepackage{amsmath}
\usepackage{amsfonts}
\usepackage{indentfirst}
\usepackage[dvips]{graphicx}

\newcommand{\bea}{\begin{eqnarray}}
\newcommand{\eea}{\end{eqnarray}}
\newcommand{\be}{\begin{equation}}
\newcommand{\ee}{\end{equation}}

\newcommand{\Z}{{\mathbb Z}}

\newcommand{\C}{{\mathbb C}}

\def\Tr{{\rm Tr}}

\def\G{\Gamma}

\newcommand{\cZ}{{\cal Z }}

\begin{document}

\sloppy

% TITLE PAGE

\begin{flushright}
\begin{tabular}{l}
CALT-68-2806 \\

\\ [.3in]
\end{tabular}
\end{flushright}

\begin{center}
\Large{ \bf Wall-crossing, open BPS counting and matrix models}
\end{center}

\begin{center}

\bigskip

\bigskip 

Piotr Su{\l}kowski\footnote{On leave from University of Amsterdam and So{\l}tan Institute for Nuclear Studies, Poland.}

\bigskip

\bigskip

\medskip 

\emph{California Institute of Technology, Pasadena, CA 91125, USA} \\ [1mm]

%\emph{psulkows@theory.caltech.edu}

\bigskip

\smallskip
 \vskip .4in \centerline{\bf Abstract}
\smallskip

\end{center}

We consider wall-crossing phenomena associated to the counting of D2-branes attached to D4-branes wrapping lagrangian cycles in Calabi-Yau manifolds, both from M-theory and matrix model perspective. Firstly, from M-theory viewpoint, we review that open BPS generating functions in various chambers are given by a restriction of the modulus square of the open topological string partition functions. Secondly, we show that these BPS generating functions can be identified with integrands of matrix models, which naturally arise in the free fermion formulation of corresponding crystal models. A parameter specifying a choice of an open BPS chamber has a natural, geometric interpretation in the crystal model. These results extend previously known relations between open topological string amplitudes and matrix models to include chamber dependence.

%*******************************************************
%*******************************************************
%*******************************************************

\newpage 

\section{Introduction}

Recently much progress has been achieved in, at first sight distinct, fields of topological strings, BPS counting, and matrix models. Also non-trivial connections between these fields have been understood. Relations between topological strings, either closed or open, and BPS counting have been known since the works of Gopakumar-Vafa and Ooguri-Vafa \cite{GV-I,GV-II,OV99}. However these relations gained new interests in view of recent results concerning stability of BPS states and associated wall-crossing phenomena \cite{DenefMoore,KS}. Among various systems undergoing these phenomena, it is especially advantageous to understand details of those which are exactly solvable. One class of such systems, which we also analyze in this paper, involves string theory on toric Calabi-Yau manifolds (more precisely, those which contain no compact four-cycles, as we will explain in what follows). This class has been studied both from physical \cite{JaffMoore,ChuangJaff,RefMotQ,OY1,ps-pyramid} and mathematical \cite{Szendroi,YoungBryan,NagaoNakajima,Nagao-strip,NagaoVO} points of view. Physically it concerns the counting of bound states of D0 and D2 branes, wrapping cycles of such toric manifolds, to a single D6-brane. In what follows we refer to such bound states as closed BPS states, and dependence of their generating functions on the moduli of the manifold is a manifestation of the wall-crossing. These results and their relation to topological string theory were explained from the M-theory perspective in \cite{WallM}. 

Another interesting series of developments relates topological string theory to matrix models. Such ideas date back to the Dijkgraaf-Vafa conjecture \cite{DV} on one hand, and on the other to the fermionic interpretation of topological strings on toric manifolds \cite{adkmv}.
Recently those connections gained new impact due to Eynard-Orantin solution of matrix models \cite{EO} and the idea of remodeling the B-model topological strings along such lines \cite{BKMP}. 
%Direct connections between topological strings and matrix models were also put forward in \cite{}

The above relations motivate a question if there is some direct relation between BPS counting and matrix models, which would take into account wall-crossing phenomena. The affirmative answer to this question was given in \cite{OSY}, where a construction of relevant matrix models was provided, based on crystal melting \cite{OY1,ps-pyramid,NagaoVO} as well as non-intersecting paths \cite{EynardTASEP} interpretations of BPS invariants. 

The results of \cite{OSY} were derived from the viewpoint of relations to closed topological strings. On the other hand, the appearance of matrix models in \cite{DV} or \cite{adkmv} was inherently related to topological branes and open topological strings. In particular, potentials of these matrix models were related to brane amplitudes, which one might think of as being integrated out to provide closed string amplitudes. In this paper we reveal similar connection between open topological strings and matrix models arising in the context of BPS counting. Our strategy is as follows. From string perspective we consider a similar system of D6-D2-D0 branes as above, which in addition includes D4-brane wrapping a lagrangian cycle in Calabi-Yau. In this context we are interested in generating functions of D2-branes with boundaries ending on this additional D4-brane, which we refer to as open BPS states. Firstly, we review that generating functions of these open BPS states can also be expressed in terms of open topological string amplitudes \cite{DSV} (these results were found independently in \cite{AY-owc}), and stress modular properties of these generating functions in a certain chamber. Then we construct matrix models, generalizing those of \cite{OSY}, whose integrands can be identified with generating functions of such open BPS states. The matrix integral itself relates these open BPS generating functions in some geometry $X$, to the closed topological string generating functions in more general geometry $Y$, which is a reminiscent of integrating brane amplitudes in the pure topological string context \cite{adkmv}. Among the others, this viewpoint sheds some light on the appearance of a non-trivial prefactor in \cite{OSY}.

It would be interesting to extend our results in various directions. A generalization to the refined BPS counting is presented in \cite{refine}. It would be interesting to include more general open BPS states (e.g. associated to several branes), understand framing dependence in general chamber, understand better a role of the more general geometry $Y$, consider  more general geometries and crystal models such as those analyzed in \cite{QuiversCrystals}, etc. We note that some other approaches to wall-crossing phenomena for open BPS states were also considered in literature. Related constructions of matrix models representing counting of closed BPS states in commutative and non-commutative chambers were introduced in \cite{SzaboTierz-DT}. Our results identify a chamber for which some particular BPS states were introduced in \cite{Nagao-open,NagaoYamazaki}. BPS generating functions similar to ours have been found in a system of closed D4-D2-D0 branes in \cite{D4D2D0}. Relation between BPS counting and lagrangian branes was also analyzed in \cite{cv-wallcross}. In the context of open topological string theory and the commutative chamber, related connections between branes, crystals and matrix models were discussed in \cite{va-sa,ps,ps-phd,bubblingCY}.

The plan of the paper is as follows. In section \ref{sec-Mtheory} we explain how generating functions of open BPS states arise from M-theory perspective \cite{DSV}. In section \ref{sec-matrix} we discuss how to encode generating functions of open BPS states in matrix models generalizing those of \cite{OSY}. In section \ref{sec-examples} we present how this identification works in several examples of toric manifolds without compact four-cycles, and discuss other aspects of our results.

%*********************************************************************
%*********************************************************************

\section{From closed to open BPS counting}      \label{sec-Mtheory}

In this section we review the relation between BPS counting, M-theory and topological strings. Starting from the counting closed BPS states \cite{WallM}, we review how to extend it to the case of open BPS states \cite{DSV} (see also \cite{AY-owc}). 

To start with, we briefly recall the results of \cite{WallM}, which considered a system of D2 and D0-branes bound to a single D6-brane in type IIA string theory. When this system is lifted to M-theory on $S^1$ following \cite{DVV-Mduality}, the D6-brane transforms into a geometric background of a Taub-NUT space with unit charge, which extends in directions transverse to the original D6-brane. This Taub-NUT space is a circle $S^1_{TN}$ fibration over $\mathbb{R}^3$, with $S^1_{TN}$ of fixed radius $R$ at infinity and shrinking to a point in the location of the D6-brane. From this perspective the counting of original bound states to the D6-brane is reinterpreted as the counting of BPS states of M2-branes in the Taub-NUT space. When the radius $R$ grows to infinity the counting does not change and ultimately can be reinterpreted in terms of a gas of free, non-interacting particles in $\mathbb{R}^5$, as long as the following two conditions are satisfied. Firstly, to avoid creation of string states arising from M5-branes wrapping four-cycles in Calabi-Yau, we simply restrict considerations to manifolds without compact four-cycles. Secondly, the moduli of the Calabi-Yau have to be tuned so that M2-branes wrapped in various ways have aligned central charges. This is achieved by considering vanishing K{\"a}hler parameters of the Calabi-Yau space, and to avoid generation of massless states, non-trivial fluxes of the M-theory three-form field through the two-cycles of the Calabi-Yau and $S^1_{TN}$ should be turned on. In type IIA this results in the $B$-field flux $B$ through two-cycles of Calabi-Yau. For a state arising from D2-brane wrapping a class $\beta$ the central charge then reads
\be
Z(l,\beta) = \frac{1}{R}(l + B\cdot \beta),         \label{Zcentral}
\ee
where $l$ counts the D0-brane charge, which is taken positive to preserve the same supersymmetry. 

Under the above conditions, the counting of D6-D2-D0 bound states is reinterpreted in terms of a gas of particles arising from M2-branes wrapped on cycles $\beta$. The excitations of these particles in $\mathbb{R}^4$, parametrized by two complex variables $z_1,z_2$, are accounted for by the modes of the holomorphic field 
\be
\Phi(z_1,z_2) = \sum_{l_1,l_2} \alpha_{l_1,l_2} z_1^{l_1} z_2^{l_2}.    \label{phi-z1z2}
\ee
Decomposing the isometry group of $\mathbb{R}^4$ as $SO(4)=SU(2)\times SU(2)'$ there are $N_{\beta}^{m,m'}$ five-dimensional BPS states of intrinsic spin $(m,m')$. We are interested in their net number arising from tracing over $SU(2)'$ spins
$$
N_{\beta}^m = \sum_{m'} (-1)^{m'} N_{\beta}^{m,m'}.
$$
The total angular momentum of a given state contributing to the index is $l=l_1+l_2+m$. Finally, in a chamber specified by the moduli $R$ and $B$, the invariant degeneracies can be expressed as the trace over the corresponding Fock space
\bea
\cZ_{BPS} & = & \Big( \textrm{Tr}_{Fock} q^{Q_0} Q^{Q_2} \Big)\, |_{chamber} = \nonumber \\
& = & \prod_{\beta,m} \prod_{l_1+l_2 = l} (1-q^{l_1+l_2+m} Q^{\beta})^{N_{\beta}^{m}} \, | _{chamber} \nonumber \\
& = & \prod_{\beta,m} \prod_{l=1}^{\infty} (1-q^{l+m} Q^{\beta})^{l N_{\beta}^{m}} \, | _{chamber},     \label{cZ-chamber}
\eea
where the subscript $chamber$ denotes restriction to those factors in the above product, which represent states which are mutually BPS
\be
Z(l,\beta)  > 0 \qquad \qquad \Leftrightarrow \qquad \qquad q^{l+m} Q^{\beta} < 1.       \label{Zpositive}
\ee
As usual, $Q=e^{-t}$ and $q=e^{-g_s}$ above encode respectively the K{\"a}hler class $t$ and the string coupling $g_s$ (we again stress that here we consider a particular class of BPS states with non-zero $B$-field, and vanishing real component of $t$). 
%Note that the product over $\beta$ runs over 
%both positive and negative classes, so that both M2 
%and anti-M2-branes contribute to the index as long 
%as the condition (\ref{Zpositive}) is satisfied.
We note that if we would restrict products in the formula (\ref{cZ-chamber}) to factors with only positive $\beta$ we would get (up to possibly some factor of MacMahon function) the Gopakumar-Vafa representation of the topological string amplitude. With all negative and positive values of $\beta$ we would get modulus square of the topological string partition function. Therefore the upshot of \cite{WallM} is that in general the above BPS generating function can be expressed in terms of the closed topological string partition function
\be
\cZ_{BPS} = \cZ_{top}(Q) \cZ_{top}(Q^{-1}) |_{chamber},     \label{ZBPS-Ztop}
\ee
where chamber restriction is to be understood as picking up only those factors in Gopakumar-Vafa product representation of $\cZ_{top}$ for which (\ref{Zpositive}) is satisfied. In this context we will often refer to the choice of a chamber as a \emph{closed BPS chamber}. The (instanton part of the) closed topological string partition function entering the above expression is given by \cite{GV-I,GV-II}
$$
\cZ_{top}(Q) = M(q)^{\chi/2} \prod_{l=1}^{\infty} \prod_{\beta>0, m} (1 - Q^{\beta} q^{m+l})^{l N^m_{\beta}},
$$
where $M(q) = \prod_l (1-q^l)^{-l}$ is the MacMahon function and $\chi$ is the Euler characteristic of the Calabi-Yau manifold. 

\bigskip

The above structure can be generalized by including in the initial D6-D2-D0 configuration additional D4-branes wrapping lagrangian cycles in the internal Calabi-Yau manifold and extending in two space-time dimensions \cite{OV99,AV2000}. For simplicity we consider a system with a single D4-brane wrapping a lagrangian cycle. There are now additional BPS states in two dimensions arising from open D2-branes ending on these D4-branes. Their net degeneracies $N_{s,\beta,\gamma}$ are characterized, firstly, by the $SO(2)$ spin $s$ whose origin is most clearly seen from the M-theory perspective \cite{OV99,LMV2000}. Secondly, they depend on two-cycles $\beta$ wrapped by open M2-branes, as well as one-cycles $\gamma$ on which these M2-branes can end. (In case of $N$ D4-branes wrapping the same lagrangian cycle, these states would additionally arise in representations $R$ of $U(N)$ \cite{OV99}. In case of a single brane this reduces to $U(1)$, and such a dependence can be reabsorbed into a parameter specifying a choice of $\gamma$.)

Lifting this system to M-theory we obtain a background of $TN_1 \, \times \,  \textrm{Calabi-Yau} \, \times \, S^1$, with the additional D4-brane promoted to M5-brane. This M5-brane wraps the lagrangian $L$ inside Calabi-Yau, the time circle $S^1$, and $\mathbb{R}_+ \times S^1_{TN}$ inside the Taub-NUT space. In particular it wraps a torus $T^2 = S^1_{TN} \times S^1$, and therefore we expect to find interesting modular properties of the BPS counting functions. As we will see, the modularity will be manifest in one chamber, where the open topological string amplitude will be completed to the product of $\theta$ functions. This M5-brane also breaks the $SO(4)$ spatial symmetry down to $SO(2)\times SO(2)'$. We denote the spins associated to both $SO(2)$ factors respectively by $s$ and $s'$, and the degeneracies of particles with such spins by $N^{s,s'}_{\beta,\gamma}$. Let us moreover introduce closed K{\"a}hler parameters $Q=e^{-t}$, as well as open ones related to discs wrapped by M2-branes $z=e^{-d}$. The real and imaginary parts of $t$ encode respectively the sizes of two-cycles $\beta$ and the value of the $B$-field through them. The real and imaginary parts of $d$ encode respectively sizes of the discs and holonomies of the gauge fields around them. Similarly as in the closed string case, to get non-trivial ensemble of mutually supersymmetric states, we set the real parts of $t$ and $d$ to zero, and consider non-trivial imaginary parts.

From the M-theory perspective we are interested in counting the net degeneracies of M2-branes ending on this M5-brane
$$
N_{s,\beta,\gamma} = \sum_{s'} (-1)^{s'} N^{s,s'}_{\beta,\gamma}.
$$ 
In the remaining three-dimensional space, in the $R\to\infty$ limit, the M2-branes ending on the M5-brane are represented by a gas of free particles. These particles have excitations in $\mathbb{R}^2$ which we identify with the $z_1$-plane. To each such BPS particle, similarly as in the closed string case \cite{WallM,DVV-Mduality}, we can associate a holomorphic field
\be
\Phi(z_1) = \sum_{l} \alpha_{l} z_1^{l}  \label{phi-z1}.
\ee 
The modes of this field create states with the intrinsic spin $s$ and the orbital momentum $l$ in the $\mathbb{R}^2$ plane. The derivation of the BPS degeneracies relies on the identification of this total momentum $s+l$ in the $R\to \infty$ limit, with the Kaluza-Klein modes associated to the rotations along $S^1_{TN}$ for the finite $R$, analogously as in the five-dimensional case discussed in \cite{DVV-Mduality,connect4d5d}.  

The BPS generating functions we are after are given by a trace over the Fock space built by the oscillators of the second quantized field $\Phi(z_1)$, and restricted to the states which are mutually supersymmetric. In such a trace each oscillator from (\ref{phi-z1}) gives rise to one factor of the form $(1 - q^{s+l-1/2} Q^{\beta} z^{\gamma} )^{\pm 1}$, where the exponent $\pm 1$ corresponds to the bosonic or fermionic character of the top component of the BPS state. This can be expressed in the CFT language, with D0-branes coupled to $L_0$ and D2-branes coupled to the currents $J_{0}^{\beta}$ and $J_0^{\gamma}$, so that we get
\bea
\cZ^{open}_{BPS} & = & \Big( \Tr \, _{Fock} \, q^{L_0} Q^{J_0^{\beta}}z^{J_0^{\gamma}} \Big)|_{chamber}  = \nonumber \\
& = & \prod_{s,\beta,\gamma} \prod_{l=1}^{\infty} (1 - q^{s+l-1/2} Q^{\beta} z^{\gamma} )^{N_{s,\beta,\gamma}}\, |_{chamber}, \label{Zbps-open}
\eea
where the product is over either both positive or both negative $(\beta,\gamma)$. The parameters $q,Q$ and $z$ specify the chamber structure: the restriction to a given $chamber$ is implemented by imposing the condition on a central charge, analogous to (\ref{Zpositive}), 
\be
q^{s+l-1/2} Q^{\beta} z^{\gamma}  <  1.         \label{Zopen}
%Z(n,\beta,\gamma) = \frac{1}{R}\big( n + B\cdot(\beta+\gamma) \big) > 0,
\ee
More precisely, this condition specifies a choice of both \emph{closed} and \emph{open} chambers. The walls of marginal stability between chambers correspond to subspaces where, for some oscillator, the above product becomes 1, and then the contribution from such an oscillator drops out from the BPS generating function.
We note that in (\ref{Zbps-open}) there is a product only over positive $l$, which from CFT perspective corresponds to the restriction to states with positive energy $L_0 > 0$, or equivalently $q^{L_0}<1$. From this viewpoint, the condition (\ref{Zopen}) can be understood similarly, as an additional restriction to states with the redefined energy $\widetilde{L}_0 >0$.

Similarly as in the closed string case, the above degeneracies can be related to open topological string amplitudes.
It was shown in \cite{OV99} that the open topological string amplitude can be written as
$$
\cZ_{top}^{open} = \exp\Big( \sum_{n=1}^{\infty} \sum_{s} \sum_{\beta,\gamma>0}  N_{s,\beta,\gamma} \frac{q^{n s} Q^{n \beta} z^{n\gamma} }{n(q^{n/2}-q^{-n/2})}   \Big)   ,  
$$
with integral invariants $N_{s,\beta,\gamma}$. In case of $N$ D4-branes wrapping a lagrangian cycle this structure is more involved, as the states in $\mathbb{R}^3$ arise also in representations of $U(N)$ \cite{OV99}, and therefore the factor $z^{n\gamma}$ is replaced by the sum $\sum_R \Tr_R V^n$ of traces in all possible representations $R$ of this $U(N)$ of the matrix $V$ encoding holonomies of the gauge fields around discs. This structure can be refined even further, by considering several distinct lagrangian submanifolds wrapped by arbitrary number of D4-branes.

The above formula for this open string partition function, as also stressed in \cite{cv-wallcross}, has essentially a structure of the quantum dilogarithm
\be
L(z,q) = \exp\Big(\sum_{n>0} \frac{z^n}{n(q^{n/2} - q^{-n/2})} \Big) = \prod_{n=1}^{\infty} (1 - z q^{n-1/2}),    \label{qdilog}
\ee
and therefore can be written in the product form 
$$
\cZ_{top}^{open}(Q,z) = \prod_{s} \prod_{\beta,\gamma>0} \prod_{n=1}^{\infty} \Big(1 - Q^{\beta}z^{\gamma} q^{s+n-1/2}  \Big)^{N_{s,\beta,\gamma}}.
$$
Comparing with (\ref{Zbps-open}) we conclude that the BPS counting functions take form of the modulus square of the open topological string amplitude
\be
\cZ^{open}_{BPS} = \cZ_{top}^{open}(Q,z) \cZ_{top}^{open}(Q^{-1},z^{-1}) \, |_{chamber}.           \label{Zbps-Ztop2}
\ee

In particular, in the extreme chamber corresponding to Im$\,t$, Im$\, d\to 0$, the trace is performed over the full Fock space and yields the modulus square of the open topological string partition function. In this case the quantum dilogarithms are completed, via the Jacobi triple product identity, to the modular function $\theta_3/\eta$, and therefore the overall BPS generating function is also modular and expressed as a product of such functions.

%*********************************************************************
%*********************************************************************

\section{Generating functions from matrix models}      \label{sec-matrix}

In this section we show that generating functions of open BPS states found in (\ref{Zbps-open}) are encoded in the structure of suitably constructed matrix models. Matrix models encoding degeneracies of closed BPS states have been found in \cite{OSY}. Here we show how to  generalize this picture to capture the counting of open BPS states. More precisely, we find that generating functions of open states found in (\ref{Zbps-open}) can be identified with integrands $e^{-\frac{1}{g_s} V(z)}$ of matrix models constructed from the associated crystal models.

\begin{figure}[htb]
\begin{center}
\includegraphics[width=0.8\textwidth]{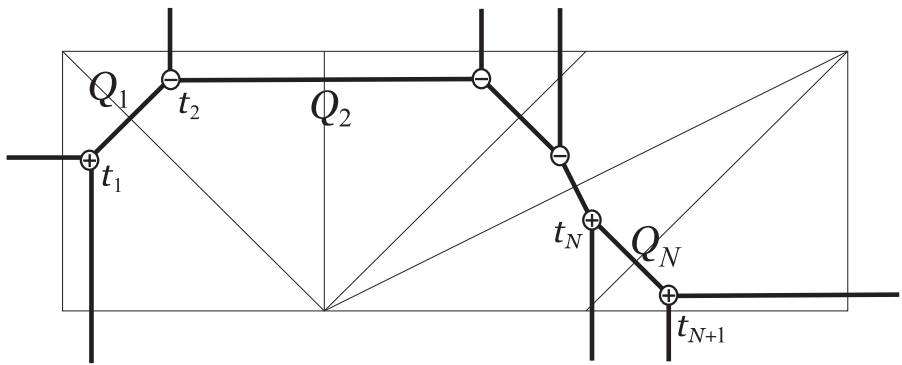} 
\begin{quote}
\caption{\emph{Toric geometries without compact four-cycles which we consider, with K{\"a}hler parameters $Q_i$ and vertices of type $t_i=\pm 1$, chosen so that $t_it_{i+1}=\pm 1$ respectively for $\C^3/\Z_2$-like and conifold-like local neighborhood of $Q_i$. }} \label{fig-strip}
\end{quote}
\end{center}
\end{figure}

To start with we recall one strategy presented in \cite{OSY}, which is based on writing generating functions of closed BPS states as fermionic correlators \cite{ps-pyramid}. Following conventions of \cite{ps-pyramid}, these correlators read
\be
Z_n \equiv \langle \Omega_+ | \overline{W}^n |\Omega_- \rangle = \cZ_{BPS},     \label{ZnW}
\ee
where $|\Omega_{\pm}\rangle$ are Bogoliubov states of the form
$$
| \Omega_- \rangle = \overline{A}_-(1) \overline{A}_-(1) \overline{A}_-(1) \ldots |0\rangle % = A_-(1) A_-(q) A_-(q^2) \ldots |0\rangle,
$$
and $\langle \Omega_+|$ are defined analogously. The structure of a given toric manifold, shown in figure \ref{fig-strip}, is encoded in operators $\overline{A}_{\pm}(1)$, which are given as products of vertex operators\footnote{In our conventions, these vertex operators satisfy the relations: $\G^{t_i}_+(x)\G^{t_j}_-(y) = (1 - t_i t_j)^{-t_i t_j} \G^{t_j}_-(y) \G^{t_i}_+(x)$, $\G^{t_i}_+(x) \widehat{Q}_k = \widehat{Q}_k \G^{t_i}_+(x q_k)$, $\widehat{Q}_k \G^{t_i}_-(x) = \G^{t_i}_-(x q_k) \widehat{Q}_k$. We also denote $\G^{(+1)}_{\pm}\equiv\G_{\pm}$ and $\G^{(-1)}_{\pm}\equiv\G'_{\pm}$.} $\G_{\pm}^{t_i}$ associated to each vertex (of type $t_i=\pm 1$, so that $t_it_{i+1}=\pm 1$ respectively for $\C^3/\Z_2$-like and conifold-like local neighborhood of $Q_i$) in a toric diagram, and weighted by $\widehat{Q}_i$ (related to the coloring $q_i$ of a given partition $\widehat{Q}_i|\lambda\rangle = q_i^{|\lambda|}|\lambda\rangle$, and encoding appropriate K{\"a}hler parameter $Q_i$)
\be
\overline{A}_{\pm}(x) = \G_{\pm}^{t_1} (x) \widehat{Q}_1 \G_{\pm}^{t_2} (x) \widehat{Q}_2 \cdots \G_{\pm}^{t_N} (x) \widehat{Q}_N \G_{\pm}^{t_{N+1}} (x) \widehat{Q}_0.                  \label{Apm}
\ee
Operators $\overline{W}^n$ encode information about a given chamber, labeled by $n$ (which is in general $N$-component vector). While most general choice of chamber, for arbitrary manifold without compact four-cycles, requires considering quite involved $\overline{W}^n$ operators \cite{NagaoVO}, for simplicity in what follows we focus on a simple class of examples where only a single K{\"a}hler parameter undergoes wall-crossing. In this case, the operator $\overline{W}^n$ is indeed the $n$'th power of $\overline{W}$ (and the integer $n$ denotes the number of walls between the given and the non-commutative chamber), and its structure, similar to $\overline{A}_{\pm}(1)$, we write down explicitly in particular examples below. 

Let us illustrate how the above formalism works in the simplest example of $\C^3$ geometry. A toric diagram in this case consists of a single toric vertex, by convention chosen to be of $t_1=+1$ type (so that a general strip geometry in fig. \ref{fig-strip} reduces to a single, left-most vertex). Therefore the operators (\ref{Apm}) take a simple form $\overline{A}_{\pm}(x) = \G_{\pm}(x)\widehat{Q}_0$, and the state associated to the manifold takes form
$$
|\Omega_-\rangle = \G_{-}(1)\widehat{Q}_0 \G_{-}(1)\widehat{Q}_0 \ldots |0\rangle = \prod_{k=0}^{\infty} \G_-(q^k) |0\rangle,
$$
and similarly for $\langle\Omega_+|$. Here we commuted all $\widehat{Q}_0$ operators to the right using relations given in the footnote below, and we identified $q=q_0$ as the eigenvalue of $\widehat{Q}_0$. As there are no K{\"a}hler parameters in this case, there are also no wall-crossing operators, and the only non-trivial chamber corresponds to $n=0$ in (\ref{ZnW}). Therefore the BPS generating function is given by the MacMahon function
$$
Z_{0} =  \langle \Omega_+ |\Omega_- \rangle = \langle 0|\Big(\prod_{i=1}^{\infty} \G_+(q^i) \Big)\Big(\prod_{k=0}^{\infty} \G_-(q^k) \Big) |0\rangle = \prod_{i,k=0}^{\infty} \frac{1}{1-q^{i+k+1}} = M(1),
$$
and in the computation we used commutation relations between $\G_{\pm}$ operators also presented in the footnote below. More complicated examples are explained at length in \cite{ps-pyramid}, and we will also discuss them in what follows.

A derivation of matrix models in \cite{OSY} relied on introducing into the correlator (\ref{ZnW}) the identity operator $\mathbb{I}$, represented by the complete set of states $|R\rangle\langle R|$. These states represent two-dimensional partitions. Using orthogonality relations of $U(\infty)$ characters $\chi_R$, and the fact that these characters are given in terms of Schur functions $\chi_R=s_R(\vec{z})$ for $\vec{z}=(z_1,z_2,z_3,\ldots)$, we can write
\bea
\mathbb{I} & = & \sum_R |R\rangle\langle R|  = \sum_{P,R} \delta_{P^t R^t} |P\rangle\langle R| = \nonumber \\
& = & \int \mathcal{D}U \sum_{P,R} s_{P^t}(\vec{z}) \overline{s_{R^t} (\vec{z})} |P\rangle\langle R| = \nonumber \\
& = & \int \mathcal{D}U \Big(\prod_{\alpha} \G_-'(z_{\alpha})|0\rangle  \Big)  \Big(\langle 0 | \prod_{\alpha} \G_+'(z^{-1}_{\alpha}) \Big),    \label{identity}
\eea
where $\mathcal{D}$ denotes the unitary measure whose eigenvalues representation is
$$
\mathcal{D}U=\prod_{\alpha} du_{\alpha} \,\prod_{\alpha<\beta}|z_{\alpha}-z_{\beta}|^2,\qquad \qquad z_{\alpha}=e^{iu_{\alpha}}.
$$  
Having inserted the identity operator in this form into (\ref{ZnW}) we can commute away $\G_{\pm}^{t_i}$ operators and get rid of operator expressions. This leads to a matrix model with the unitary measure
\be
Z_n = f_n(q,Q_i) \, \int \mathcal{D}U \prod_{\alpha} e^{-\frac{1}{g_s} V(z_{\alpha})},   \label{Zmatrix}
\ee
where the product over $\alpha$ represents distinct eigenvalues $z_{\alpha}$. For generic chambers we find some overall factors $f_n(q,Q_i)$. These factors, in generic chamber, arise from commutations between certain $\G_{\pm}$ components of wall-crossing operators, and $\G_{\mp}$ components of $|\Omega_{\mp}\rangle$ states, and take form of relatively simple infinite products. Importantly, these factors do not depend on parameters labeling open chambers -- in this sense matrix model indeed encodes open chamber dependence, as we discuss below. Moreover, in the non-commutative chamber these factors reduce to $f_{n=0}(q,Q_i)=1$, and they largely simplify in the commutative chamber $n\to \infty$, as will become clear in the examples presented below.

In case of closed BPS states, the above matrix model representation of $Z_n$ depends on K{\"a}hler parameters $Q_i$ encoded in the potential $V(z)$, and the choice of closed BPS chamber is specified by the number $n$ of wall-crossing operators in (\ref{ZnW}). In the context of counting open BPS states one should introduce their generating parameter, as well as specify a choice of open BPS chamber $k$. Now we present how to identify these parameters.

\begin{figure}[htb]
\begin{center}
\includegraphics[width=0.4\textwidth]{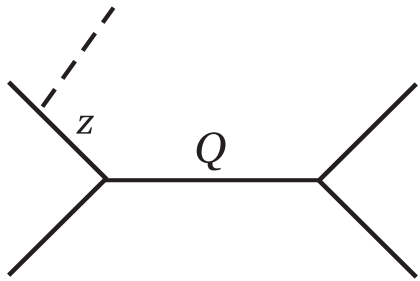} 
\begin{quote}
\caption{\emph{Brane associated to the external leg of a toric diagram (of a conifold in this particular case). Closed string parameter is denoted by $Q$ and open string parameter by $z$.}} \label{fig-conibrane}
\end{quote}
\end{center}
\end{figure}

Firstly, we claim that the open generating parameter can be identified with matrix eigenvalues $z_{\alpha}$, and a dependence on an open BPS chamber can be introduced by a more general way of insertion of the identity operator in (\ref{ZnW}). In particular, the open BPS chamber labeled by $k$ is be represented by inserting the identity at location $k$ within a string of $\overline{A}_-$ operators.
%\footnote{In \cite{OSY} the location of the identity 
%was fixed and no such dependence on $k$ arose.} 
We recall that the correlator (\ref{ZnW}) represents a pyramid crystal, in which each single-colored layer is represented by an insertion of one $\G_{\pm}$ operator (which are building blocks of $\overline{A}_{\pm}$). Therefore the insertion of the identity operator in the form (\ref{identity}) at a position $k$ represents gluing  a crystal, along the corresponding layer, from the two independent parts. This is a reminiscent of how closed topological string amplitudes are built from the open amplitudes in the topological vertex theory  \cite{adkmv}. Therefore, the above prescription leads to the following matrix model representation
\bea
Z_n & = & \langle 0 | \prod_{i=k}^{\infty}  \overline{A}_+(1) | \mathbb{I} | \prod_{j=0}^{k-1}  \overline{A}_+(1) | \overline{W}^n  |\Omega_-\rangle = \nonumber \\
& = & \int \mathcal{D}U  \ \langle 0 | \prod_{i=k+1}^{\infty}  \overline{A}_+(1) |
\prod_{\alpha} \G_-'(z_{\alpha})|0\rangle  \langle 0 | \prod_{\alpha} \G_+'(z^{-1}_{\alpha}) | \prod_{j=0}^{k}  \overline{A}_+(1) | \overline{W}^n | \Omega_-\rangle =   \nonumber   \\
& = & f^k_{n}(q,Q_i) \int \mathcal{D}U \prod_{\alpha} e^{-\frac{1}{g_s} V^{k}_{n}(z_{\alpha})},  \label{Zkn-int}
\eea
and our claim states that the open BPS generating function (\ref{Zbps-Ztop2})  can be identified with the integrand
\be
\cZ^{open}_{BPS} = 
%\cZ_{top}^{open}(\mathcal{Q},z) \cZ_{top}^{open}(\mathcal{Q}^{-1},z^{-1}) \, |_{chamber} = 
e^{-\frac{1}{g_s} V^{k}_{n}(z)},   \label{claim}
\ee
up to a simple identification of parameters (which amounts to the shift $z\to-zq^{1/2}$ (to match earlier M-theory convention with half-integer powers of $q$, to integer powers of $q$ in the fermionic formalism), as well as identification of K{\"a}hler parameters considered in M-theory derivation with $\mu_i$ introduced below). The BPS generating function in (\ref{Zbps-Ztop2}) is determined by the open topological string partition function associated to the external axis of the toric diagram, as in figure \ref{fig-conibrane}.\footnote{Similarly as in topological strings, one should be able to obtain amplitudes for branes associated to other axis by appropriate analytic continuation}
The prefactor $f^k_n$ above arises from commuting away vertex operators and it depends on $q$ and closed string parameters $Q_i$, but does not involve open generating parameters. Now the potential $V^k_n(z)$ depends on a choice of both closed and open chambers, specified by integers $n$ and $k$, and open BPS modulus is identified with matrix eigenvalue $z$. 

Secondly, we claim that the value of the above integral can be related to a more general Calabi-Yau geometry $Y$. This more general geometry involves two copies of the initial geometry  $X$ with K{\"a}hler parameters $Q_i$ and $\mu_i$ respectively (where $\mu_i$ encode information of the closed BPS chambers), as well as an additional two-cycle with K{\"a}hler parameter $q^k$. The value of the integral $\int \mathcal{D}U \prod_{\alpha} e^{-\frac{1}{g_s} V^{k}_{n}(z_{\alpha})}$ is then given by the part of the closed topological string amplitude for $Y$ which probes this additional two-cycle (i.e. contains only factors involving $q^k$). 

%*********************************************************************

Let us mention how how these results relate to the viewpoint of \cite{adkmv}. In general in string theory closed amplitudes can be obtained from integrating out open amplitudes
\be
\cZ_{closed} = \int \cZ_{open}.    \label{ZclosedOpen}
\ee
As stressed in \cite{adkmv}, topological strings are a very instructive example of such phenomena, where integrals over open degrees of freedom reduce to matrix integrals, whose potentials encode (a reduction of) the action on a brane. Precise form of such integrals depends on a nature of open degrees of freedom. In case of compact branes, the theory on a brane is identified with a holomorphic Chern-Simons theory, whose reduction results in Dijkgraaf-Vafa matrix models \cite{DV}. On the other hand, the case of non-compact branes on B-model geometries $uv + H(x,y) = 0$ involving a Riemann surface $H(x,y)=0$ leads to matrix models 
\be
\cZ_{closed} = \int \cZ_{open} \sim \int \mathcal{D}U e^{-\frac{1}{g_s} \int y(x) dx},  \label{ZclosedKS}
\ee
whose action (in the WKB approximation) is identified with the with Kodaira-Spencer field $\phi(x)=\int^x y(x)dx$. The expression for $y(x)$ can be found from the equation of the Riemann surface  $H(x,y)=0$, and the integration by parts in (\ref{ZclosedKS}) leads to the Kontsevich-like matrix models presented in \cite{adkmv}. This case of non-compact branes on geometries based on a Riemann surface is also important to us, as it can be thought of as a limiting (commutative) case of situations involving BPS counting. Our results give rise to a relation similar to (\ref{ZclosedOpen}), however valid in principle in any chamber. In our case $Z_{open}$ refers to the open BPS amplitude in an arbitrary (open and closed) chamber in the initial geometry $X$, while  $Z_{closed}$ represents (a part of) the closed topological string amplitude for the geometry $Y$ introduced above. On the other hand, in the commutative open and closed chamber $k,n\to \infty$, our matrix models have form consistent with (\ref{ZclosedKS}). 

In the next section we present how these claims are realized in several representative examples. We note that in the open chamber corresponding to $k\to\infty$, the representation of the identity operator in the first line of (\ref{Zkn-int}) detects the open BPS configuration corresponding to the open BPS chamber implied by a definition considered in \cite{Nagao-open,NagaoYamazaki}. We also note that related geometric transitions in the purely topological string perspective were discussed in \cite{bubblingCY}.

%*********************************************************************
%*********************************************************************

\section{Examples}       \label{sec-examples}

In this section we show in several examples how open BPS generating functions arise from matrix model realization of closed BPS generating functions, and present associated generalized geometries $Y$. Examples which we discuss involve $\mathbb{C}^3$ geometry with arbitrary open BPS chamber, arbitrary geometry with arbitrary open chamber and fixed (non-commutative) closed chamber, as well as conifold and $\mathbb{C}^3/\Z_2$ geometries with arbitrary open and closed chambers.

One more remark is in order here. As shown in section \ref{sec-Mtheory}, open BPS generating functions can be identified with a reduction of the modulus square of the open topological string amplitude. In fact we should notice that such open topological amplitudes are defined up to a framing ambiguity \cite{AKV2001,FramedKnots}. This shows up already in the simplest example of $\mathbb{C}^3$ geometry, which in generic framing encodes infinite number of open Gopakumar-Vafa invariants. Nonetheless, for branes in a geometry without compact four-cycles, one can always choose a framing in which amplitudes simplify, and are given by a product of a few quantum dilogarithms representing wrappings over all possible open and closed two-cycles in the manifold. Such amplitudes were explicitly computed for example in \cite{adkmv,va-sa,ps}. In case of a single brane in $\mathbb{C}^3$, in such special framing the amplitude is given by a single quantum dilogarithm (\ref{qdilog})
\be
\cZ^{open,\, \mathbb{C}^3}_{top} = L(z,q),   \label{ZtopC3}
\ee
where $z$ captures the area of a disc ending on the lagrangian cycle. Similarly, for a brane in the (external leg of the) conifold, there is a special framing in which brane amplitude is given by a ratio of two dilogarithms. In what follows we implicitly consider branes in such special framings. For general strip-like geometry with $N+1$ vertices of types $t_i$ (with convention $t_1=+1$) and K{\"a}hler parameters $Q_i$, such brane amplitude reads
\be
\cZ^{open}_{top} = \prod_{a=1}^{N+1} L\big(z (Q_1Q_2\cdots Q_{a-1}),q\big)^{t_a} \equiv L(z) \prod_{a=2}^{N+1} L\big(z (Q_1Q_2\cdots Q_{a-1}),q\big)^{t_a},   \label{Ztop}
\ee
and the corresponding open BPS generating function is
\be
\cZ^{open}_{BPS} = \prod_{a=1}^{N+1} L\big(z (Q_1Q_2\cdots Q_{a-1}),q\big)^{t_a} \, L\big(z^{-1} (Q_1Q_2\cdots Q_{a-1})^{-1},q\big)^{t_a} |_{chamber}.   \label{ZopenBPS}
\ee
It would be interesting to understand framing dependence of these amplitudes as well.

%*********************************************************************

\subsection{$\mathbb{C}^3$}

As the first example we consider $\mathbb{C}^3$ geometry and show how open BPS generating functions, derived from M-theory viewpoint in section \ref{sec-Mtheory}, appear in matrix models encoding closed amplitudes, discussed in section \ref{sec-matrix}. 

As already mentioned, the open topological string amplitude for a brane in $\mathbb{C}^3$ is given by the quantum dilogarithm (\ref{ZtopC3}). From the stability condition for the central charge (\ref{Zopen}) and the general formula (\ref{Zbps-Ztop2}), we find, in the open chamber labeled by $k$, the following set of open BPS generating functions
\be
\cZ^{open}_{k} = \prod_{i=1}^{\infty} (1 - z q^{i-1/2}) \prod_{j>k}^{\infty} (1 - z^{-1} q^{j-1/2}).    \label{ZbpsC3-1}
\ee
Let us note that the chamber for $k=0$ is a special one, in which the generating function
$$
Z_0 = q^{1/24}\frac{\theta_3(z,q)}{\eta(q)},
$$
is indeed (up to the overall $q^{1/24}$) a modular form, as explained in section \ref{sec-Mtheory}. On the other hand, for $k\to\infty$, the generating function  $Z_{k\to\infty}$ reduces to the open topological string amplitude (\ref{ZtopC3}). 

\begin{figure}[htb]
\begin{center}
\includegraphics[width=0.35\textwidth]{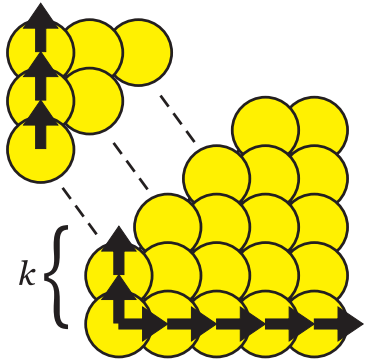} 
\begin{quote}
\caption{\emph{Factorization of $\mathbb{C}^3$ crystal which leads to open BPS generating functions. The size $k$ encodes the open BPS chamber.}} \label{fig-C3split2}
\end{quote}
\end{center}
\end{figure}

\begin{figure}[htb]
\begin{center}
\includegraphics[width=0.35\textwidth]{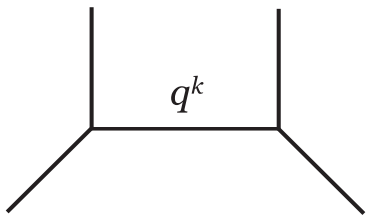} 
\begin{quote}
\caption{\emph{Effective geometry detected by $Z_{matrix}$ in (\ref{ZmatrixC3}).}} \label{fig-C3Z2}
\end{quote}
\end{center}
\end{figure}

Let us present how the result (\ref{ZbpsC3-1}) arises from the matrix model viewpoint, following conventions presented in section \ref{sec-matrix}. In this case $\overline{A}_+(1) = \G_+(1) \widehat{Q}$ and the geometry of $\mathbb{C}^3$ is encoded in the state
$$
|\Omega_-\rangle = \prod_{i=1}^{\infty} \G_-(q^i)|0\rangle,
$$
and similarly for $\langle\Omega_+|$. There is a single closed string chamber in which the generating function $Z=\langle\Omega_+|\Omega_-\rangle = M(1)$ is given by the MacMahon function $M(1)$, where we denote
$$
M(x) = \prod_{j=1}^{\infty} \frac{1}{(1-x q^j)^j}.
$$
According to our proposition, we now insert the operator $\mathbb{I}$ at the location $k$. While it does not change the total value of matrix integral, the explicit dependence on $k$ enters the potential and we get:
\bea
Z & = & M(1) = \langle 0 | \prod_{i=k}^{\infty} \overline{A}_+(1) | \mathbb{I} |  \prod_{j=0}^{k-1} \overline{A}_+(1) | \Omega_-\rangle = \nonumber  \\
& = & f^k(q) Z_{matrix},    \label{C3open-matrix}
\eea 
where
$$
Z_{matrix} = \int \mathcal{D}U \prod_{\alpha} \prod_{j=1}^{\infty} (1 + z_{\alpha}q^j) \prod_{i=k}^{\infty} (1 + z_{\alpha}^{-1}q^j) ,
$$
and
$$
f^k(q) = \prod_{i=1}^{k} \prod_{j=0}^{\infty} \frac{1}{1-q^{i+j}} = \frac{M(1)}{M(q^k)}.
$$
Matrix model integrand indeed reproduces open BPS generating function (\ref{ZbpsC3-1}) (up to a redefinition $z\to -z q^{1/2}$) in a chamber labeled by $k$. From the viewpoint of open BPS states, the prefactor $f^k(q)$ may be viewed as an ingredient necessary to provide the required form of the matrix model potential. The value of the matrix integral itself,
\be
Z_{matrix} = M(q^k),    \label{ZmatrixC3}
\ee
can be identified with a part (sensitive to $q^k$) of the closed topological string geometry $Y=\C^3/\Z_2$, with $\mathbb{P}^1$ resolved to size $q^k$, as shown in figure \ref{fig-C3Z2}. 

We note that in the limit $k\to \infty$ the potential, to the leading order, is given by the dilogarithm
$$
L(z,q) \sim e^{-\frac{1}{g_s} \textrm{Li}_2(z)}.
$$
This is consistent with (\ref{ZclosedKS}). Indeed, from the equation of the mirror surface $H(x,y)= -e^x + e^{-y} - 1 = 0$ we find 
$$
e^{-\frac{1}{g_s} \int y(x) dx} = e^{-\frac{1}{g_s}\textrm{Li}_2(e^{x+i\pi})},
$$
which reproduces to the leading order the potential arising from the quantum dilogarithm.

%*********************************************************************

\subsection{Arbitrary geometry}

Now we consider an arbitrary strip-like geometry, with arbitrary open BPS chamber, and with all closed chambers fixed to the non-commutative value $n=0$. The correlators (\ref{Zkn-int}) result in 
$$
Z_{n=0} = f^k_{n=0} Z_{matrix},
$$
where, in terms of $\mu_i=\frac{1}{Q_i}=(t_it_{i+1})\frac{1}{q_i}$, we find
$$
Z_{matrix} =  \int \mathcal{D}U \prod_{\alpha} \prod_{a=1}^{N+1} \prod_{j=0}^{\infty} \big(1 + z_{\alpha}q^{j+1} (\mu_1\mu_2\cdots \mu_{a-1}) \big)^{t_a} \big(1 + \frac{q^{j+k} }{z_{\alpha}\mu_1\mu_2\cdots \mu_{a-1}} \big)^{t_a}. 
$$
This agrees with open BPS generating function (\ref{ZopenBPS}) in $n=0$ chamber, if one redefines $z\to -z q^{1/2}$ as above, and identifies K{\"a}hler parameters used in M-theory derivation of (\ref{ZopenBPS}) with $\mu_i$. Indeed, considering the open chamber labeled by $k$ amounts to a chamber restriction in (\ref{ZopenBPS}) which is manifested by including a factor $q^k$ in the argument of all dilogarithms involving $z^{-1}$. 

\begin{figure}[htb]
\begin{center}
\includegraphics[width=0.7\textwidth]{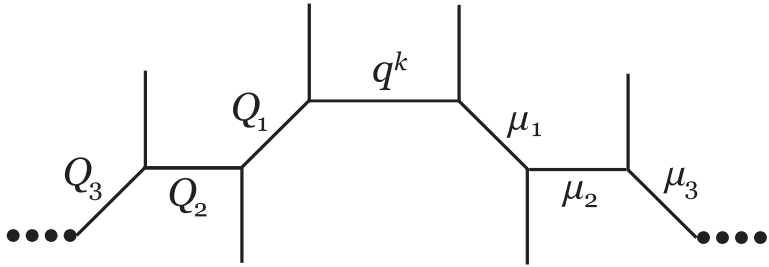} 
\begin{quote}
\caption{\emph{Effective geometry detected by $Z_{matrix}$ in (\ref{Zmatrix-general}).}} \label{fig-C3Z2general}
\end{quote}
\end{center}
\end{figure}

The factor $f^k_{n=0}$ is given as the ratio of $Z_{n=0}$, and $Z_{n=0}$ with all arguments of MacMahon functions shifted by $q^k$. This implies that
\be
Z_{matrix} = M(q^k)^{N+1} \prod_{1\leq a<b\leq N+1} M(q^k Q_a\cdots Q_{b-1})^{t_a t_b} M(q^k \mu_a\cdots \mu_{b-1})^{t_a t_b}.     \label{Zmatrix-general}
\ee
This result can be interpreted as $q^k$-sensitive part of the closed topological string amplitude for a more general geometry $Y$, which includes two copies of the initial geometry $X$ with K{\"a}hler parameters $Q_i$ and $\mu_i=Q_i^{-1}$ respectively, as well as an additional two-cycle of type $\mathbb{C}^3/\Z_2$, see figure \ref{fig-C3Z2general}. 

%*********************************************************************

\subsection{Conifold}

We repeat the above analysis for the case of a brane associated to the external leg of a toric diagram. With appropriate choice of the framing its amplitude reads
$$
\cZ_{top}^{open} = \frac{L(z,q)}{L(zQ,q)},    % Z_{top}^{int} = L(z,q) L(Q/z,q),
$$
In the context of open BPS counting, this again leads to the modular generating function in the non-commutative chamber $n=k=0$. 
%This statement also extends to the case of several branes. For example, the amplitude with% two branes (on external and internal leg) considered at the same time reads
%$$
%\cZ_{top}^{int,ext} = \frac{1}{1-z_1z_2} \frac{L(z_1,q)L(Q/z_1,q)L(z_2,q)}{L(Qz_2,q)},
%$$
%and in the extreme chamber the amplitude is modular and expressed as a product of function%s $\theta_3/\eta$.

More generally, let us consider open BPS counting associated to such a brane, with closed chamber labeled by $n$, and open chamber labeled by $k$. The analysis of the condition (\ref{Zopen}) leads, after the shift  $z\to -z q^{1/2}$, to a general generating function of open BPS states
\be
\cZ^{open, \, k}_{n} = |\cZ^{open}_{top}|^2_{\ chamber} = 
\prod_{l=1}^{\infty} \frac{(1 + z q^l)(1 + z^{-1} q^{k+l-1})}{(1 + zQ q^l)(1 + z^{-1}Q^{-1} q^{n+k+l-1})}. \label{Znk-coni}
\ee

\begin{figure}[htb]
\begin{center}
\includegraphics[width=0.7\textwidth]{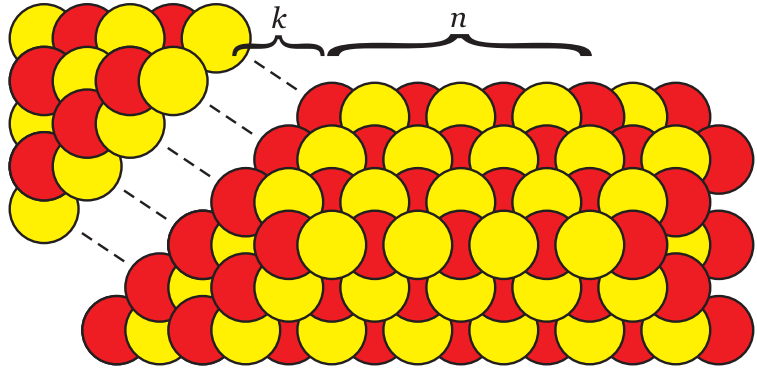} 
\begin{quote}
\caption{\emph{Factorization of the conifold pyramid which leads to open BPS generating functions. The size of the pyramid $n$ represents the closed BPS chamber, while the size $k$ encodes the open BPS chamber.}} \label{fig-coni-split}
\end{quote}
\end{center}
\end{figure}

Again we can show that it arises from matrix model viewpoint. Following conventions of section \ref{sec-matrix} and precise form of operators $\overline{A}_{\pm}(x) = \G_{\pm}(x) \widehat{Q}_1 \G'_{\pm}(x) \widehat{Q}_0$ and $\overline{W} = \G_{-}(1) \widehat{Q}_1 \G'_{+}(1) \widehat{Q}_0$ derived in \cite{ps-pyramid}, we find
\be
Z_n  =  \langle 0 | \prod_{i=k}^{\infty}  \overline{A}_+(1) | \mathbb{I} | \prod_{j=0}^{k-1}  \overline{A}_+(1) |\overline{W}^n | \Omega_-\rangle = f^{k}_{n} Z_{matrix}.
\ee
In terms of $\mu=-\frac{1}{q_1}=Q^{-1} q^n$ the matrix integral takes form
$$
Z_{matrix} =  \int \mathcal{D}U \prod_{\alpha} \prod_{j=1}^{\infty} \frac{(1 + z_{\alpha}q^{j})(1 + z_{\alpha}^{-1} q^{k+j-1}) }{(1 + z_{\alpha} \mu q^{j})
(1+  z^{-1}_{\alpha}\mu^{-1} q^{j+n+k-1})}.
$$
The integrand of this matrix model agrees with M-theory considerations (\ref{Znk-coni}) (again identifying $\mu$ with K{\"a}hler parameter  used in M-theory derivation). In the limit $n\to\infty$ followed by $\mu\to 0$ we get the answer for $\mathbb{C}^3$ given in (\ref{C3open-matrix}), as expected. On the other hand, for both $n,k\to\infty$, the integrand reduces to the open topological string amplitude given by a ratio of two quantum dilogarithms. To the leading order this ratio is is equal to a difference of two ordinary dilogarithms $V_{n,k\to\infty}\sim \textrm{Li}_2(e^{x+i\pi})-\textrm{Li}_2(\mu e^{x+i\pi})$, in agreement with the form of matrix models (\ref{ZclosedKS}) derived from the conifold geometry $H(x,y)=1+e^x + e^{-y} + \mu e^{x-y}$. 

\begin{figure}[htb]
\begin{center}
\includegraphics[width=0.45\textwidth]{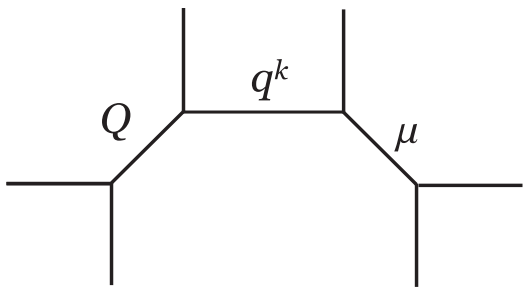} 
\begin{quote}
\caption{\emph{Effective geometry detected by $Z_{matrix}$ in (\ref{Zmatrix-coni}}).} \label{fig-C3Z2coni}
\end{quote}
\end{center}
\end{figure}

The prefactor above is found as
$$
f^{k}_{n} = M(1)^2 \frac{M(\mu q^{k}) M(Q q^{k}) }{M(\mu) M(q^{k}) M(Q) M(\mu Q q^{k})} \prod_{j=1}^{\infty}\big(1 - \mu q^j \big)^n.
$$
In consequence, the value of the matrix integral takes form
\be
Z_{matrix} = \frac{M(q^k) M(\mu Q q^k)}{M(\mu q^k) M(Q q^k)},    \label{Zmatrix-coni}
\ee
which is the $q^k$-sensitive part of the closed topological string partition function of a manifold $Y$ shown in figure \ref{fig-C3Z2coni}. As claimed above, $Y$ consists of two copies of conifolds parametrized respectively by $Q$ and $\mu$, as well as an additional cycle of size $q^k$. 

%*********************************************************************

\subsection{$\mathbb{C}^3/\mathbb{Z}_2$}

Finally we consider the resolved $\mathbb{C}^3/\mathbb{Z}_2$ singularity. In this case the topological string partition function for a brane on the external leg reads
$$
Z^{ext}_{top} = L(z,q) L(zQ,q).
$$
In consequence, BPS generating functions in a closed chamber $n$ and open chamber $k$ is (after  $z\to -z q^{1/2}$ shift)
$$
\cZ^{open,\, k}_{n} = |\cZ^{open}_{top}|^2 _{chamber} = \prod_{l=1}^{\infty} (1+z q^l)(1 + zQ q^l)(1+z^{-1} q^{k+l-1})(1+z^{-1}Q^{-1} q^{n+k+l-1}).
$$

On the other hand, from matrix model perspective, with  $\overline{A}_{\pm}(x) = \G_{\pm}(x) Q_1 \G_{\pm}(x) Q_0$  and $\overline{W} = \G_{-}(1) \widehat{Q}_1 \G_{+}(1) \widehat{Q}_0$ as discussed in \cite{ps-pyramid}, and $\mu=\frac{1}{q_1}=Q^{-1} q^n$, we obtain
\bea
Z^n_k & = & \langle 0 | \prod_{i=k}^{\infty}  \overline{A}_+(1) | \mathbb{I} | \prod_{j=0}^{k-1}  \overline{A}_+(1) | \overline{W}^n | \Omega_-\rangle = f^k_n Z_{matrix} = \nonumber \\
& = & f^{k}_{n} \int \mathcal{D}U \prod_{\alpha} \prod_{j=1}^{\infty} (1 + z_{\alpha}q^{j})(1 + z_{\alpha}\mu  q^{j})
(1 + \frac{q^{k+j-1}}{z_{\alpha}})(1+ \frac{q^{n+k+j-1}}{z_{\alpha}\mu}),
\eea
and the matrix integrand again agrees with the M-theory result (when written in terms of the argument $\mu$) above.

\begin{figure}[htb]
\begin{center}
\includegraphics[width=0.35\textwidth]{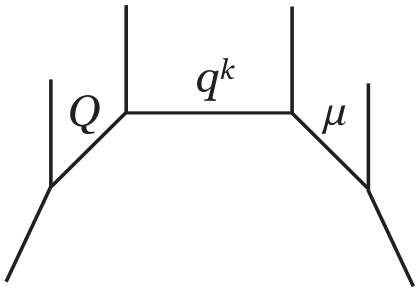} 
\begin{quote}
\caption{\emph{Effective geometry detected by $Z_{matrix}$ in (\ref{Zmatrix-triple}).}} \label{fig-C3Z2-triple}
\end{quote}
\end{center}
\end{figure}

Now the prefactor reads
$$
f^{k}_{n} = M(1)^2 \frac{M(\mu) M(Q)}{ M(\mu q^{k}) M(q^{k}) M(\mu Q q^{k}) M(Q q^{k})} \prod_{j=1}^{\infty}(1-\mu q^j)^{-n}.
$$
Therefore the matrix integral takes value
\be
Z_{matrix} =  M(q^{k}) M(\mu q^{k}) M(Q q^{k}) M(\mu Q q^{k}),    \label{Zmatrix-triple}
\ee
which is $q^k$-sensitive part of the geometry shown in figure \ref{fig-C3Z2-triple}.

%*******************************************************************
%*******************************************************************
    
\newpage

\bigskip

\begin{center}
{\bf Acknowledgments}
\end{center}

\medskip

I thank Hirosi Ooguri and Masahito Yamazaki for inspiring discussions and collaboration on related projects. This research was supported by the DOE grant DE-FG03-92ER40701FG-02 and the European Commission under the Marie-Curie International Outgoing Fellowship Programme. The contents of this publication reflect only the views of the author and not the views of the funding agencies.

%*******************************************************************
%*******************************************************************

\end{document}